# Photoacoustics Modelling using Amplitude Mode Expansion Method in a Multi-scale T-cell Resonator

S. El-Busaidy[1,2], B. Baumann[1], M. Wolff[1], L. Duggen[2]
1. Department of Mechanical and Production Engineering, Hamburg University of Applied Sciences, Hamburg, Germany
2. Mads Clausen Institute, University of Southern Denmark, Sønderborg, Denmark

**Introduction**

Photoacoustic spectroscopy (PAS) is a highly sensitive technique with the ability to detect parts-per-billion (ppb) levels of species in samples and measure optically opaque samples [1]. The technique employs modulated electromagnetic radiation to excite molecules to higher energetic states. The excited molecules relax non-radiatively through heat release to the surrounding environment. Since the radiation is modulated, the heat release is periodic. This generates pressure changes in the surrounding environment that are detected as acoustic waves using a microphone or a tuning fork [2, 3, 4].

This photoacoustic (PA) signal is usually weak and an acoustic resonator is employed for signal amplification. The electromagnetic radiation is modulated at an acoustic eigenfrequency of the resonator to excite the corresponding acoustic mode. This results in an enhanced PA signal and an increased sensitivity of the technique. Therefore, optimization of the resonator's geometry is important for maximizing the PA signal of a measurement system. Since the geometry of the resonator that results in the highest acoustic amplification is not obvious, often various resonator shapes are tested. Experimentally testing a larger number of shapes would be extremely time consuming and expensive. Therefore, numerical simulation methods are preferred.

Due to its simplicity, the transmission line model is a common method for the investigation of PA resonators [5]. However, it is a one-dimensional method and, therefore, unable to take radial and azimuthal modes into account that maybe present in a resonator. The viscothermal (VT) model is considered the most accurate numerical method for simulating PA signals [6]. The method requires the use of boundary layers which can accurately map the loss effects at the surfaces of the resonator. This is particularly important since surface losses are the dominant loss mechanism in acoustic resonators. The VT model is computationally demanding requiring a lot of memory space and simulation time.

In this article, we investigate the amplitude mode expansion (AME) model which is considered faster and computationally less demanding than the VT model [7]. The method is used to simulate the PA signal of a multi-scale T-cell resonator which consists of three interconnected cylinders as shown in Figure 1. A small *absorption cylinder* is longitudinally connected to a *cavity cylinder* which has a *resonance cylinder* perpendicularly mounted, thus forming a T-like structure. The idea behind this geometry is to press a solid state sample onto the absorption cylinder thus sealing the left end of the resonator. The upper end of the resonance cylinder is sealed by the microphone and its mounting. The sample is excited by a laser beam which enters the resonator through a window sealing the opening at the right end of the cavity cylinder.

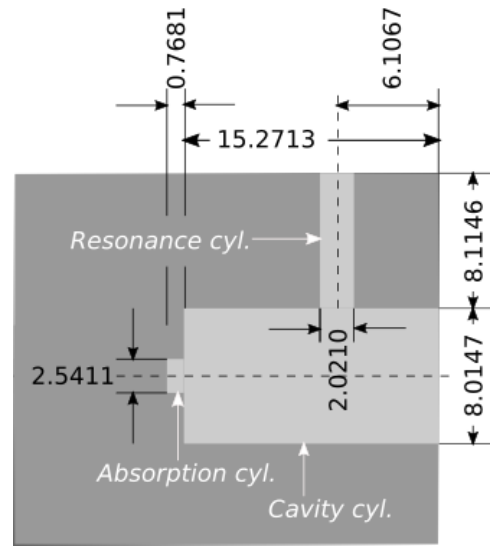

**Figure 1**: Cross-section of the T-cell resonator (light gray) showing the size of each cylinder in mm. The dimensions of the cells are the mean values obtained from a high precision measurement of the resonator [9].

Previous studies using the AME model for an investigation of a T-cell resonator were done in the ultrasound range [8]. Glière et al. compared the results of the AME model to those of the VT model and a third modeling approach in a micro-resonator by looking at a single resonance [6]. Here we will extend the study by simulating the PA signal of a macroscopic T-cell over a wide frequency range of 8 to 62 kHz.

## Theoretical background

In this section, we provide a basic description of the two methods of interest.

### AME model

The AME method is based on calculating the acoustic pressure at the location of the microphone by solving the inhomogeneous Helmholtz equation

$$\nabla^2 p(\mathbf{r},\omega) + k^2 p(\mathbf{r},\omega) = i\omega \frac{\gamma-1}{c^2} \mathcal{H}(\mathbf{r},\omega), \quad (1)$$

where $p(\mathbf{r},\omega)$ is the acoustic pressure at the measurement point $\mathbf{r}$ and modulation frequency $\omega$, $\gamma$ is the ratio of isobaric and isochoric heat capacity, $k$ is the acoustic wave number and $c$ is the speed of sound. The right-hand side of the equation describes the excitation of the acoustic waves. $\mathcal{H}(\mathbf{r},\omega)$ is the power density at modulation frequency $\omega$, obtained by applying the Fourier transform to the time dependent input $H(\mathbf{r},t)$. It is related to the radiation intensity $I$ by $\mathcal{H}(\mathbf{r},\omega) = \alpha I(\mathbf{r},\omega)$, where $\alpha$ is the absorption coefficient of the sample.

The solution of Equation 1 can be expressed as a superposition of the acoustical eigenmodes of the resonator

$$p(\mathbf{r},\omega) = \sum_j A_j(\omega) p_j(\mathbf{r}). \quad (2)$$

The modes $p_j(\mathbf{r})$ and corresponding natural frequencies $\omega_j$ are obtained by solving the homogeneous Helmholtz equation with a sound hard boundary condition

$$\nabla^2 p(\mathbf{r}) + k^2 p(\mathbf{r}) = 0, \quad (3)$$

while the amplitudes $A_j(\omega)$ are obtained using

$$A_j(\omega) = i \frac{\mathcal{A}_j \omega}{\omega^2 - \omega_j^2 + i\omega \omega_j l_j}. \quad (4)$$

$\mathcal{A}_j$ is calculated by

$$\mathcal{A}_j = \frac{\alpha(\gamma-1)}{V_c} \int_{V_c} p_j^* I \, dV, \quad (5)$$

where $V_c$ denotes the volume of the resonator and the asterisk indicates complex conjugation. The loss effects are introduced by loss factors $l_j$ in Equation 4. There are numerous loss effects that attenuate the PAS signal, however, the method only considers the surface and volume losses due to viscosity and thermal conduction. A detailed exposition can be found elsewhere [7].

### VT model

The method is based on solving the linearized Navier-Stokes equation, the continuity equation for the mass, and the energy balance equation. An equation of state is introduced so as to relate the variations in pressure, temperature and density.

Using the propagating fluid's parameters, the equations can be solved for small perturbations in the acoustic pressure, temperature and velocity vector. A detailed description of the equations can be found elsewhere [10].

## Numerical models

This section describes how the two models are realized and implemented.

### AME model

The model is implemented using a MATLAB® script. The material properties of the propagating fluid in the resonator (air in our case) are found in Table 1.

**Table 1**: Air parameters at a temperature of 20°C and a static pressure of 1013 hPa [11].

| | |
|---|---|
| Density | 1.2044 kg/m$^3$ |
| Sound velocity | 343.2 m/s |
| Viscosity | 1.814 10$^{-5}$ Pa s |
| Coefficient of heat conduction | 2.58 10$^{-2}$ W/m K |
| Specific heat capacity at constant volume | 7.1816 10$^2$ J/kg K |
| Specific heat capacity at constant pressure | 1.0054 10$^3$ J/kg K |

The geometry of the resonator is initially created and meshed. The mesh consists of swept meshes in the resonance cylinder and some sections of the cavity cylinder, while a triangular mesh was used in areas of the resonator where a swept mesh could not be applied. A structured mesh with swept meshing was preferred in order to reduce the number of mesh elements and enable faster computation.

The script accesses the Pressure Acoustics, Frequency Domain COMSOL Multiphysics® module using the COMSOL LiveLink™. The resonator walls are set as sound hard. The resonator's eigenfrequencies and eigenmodes are calculated using equation 3.

Once the eigenfrequencies and eigenmodes have been calculated, the MATLAB script receives the solutions and proceeds to calculate the loss factors and the amplitudes. The source term in Equation 5 is defined within the absorption cylinder as shown in Figure 2.

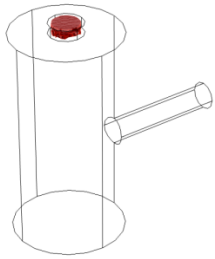 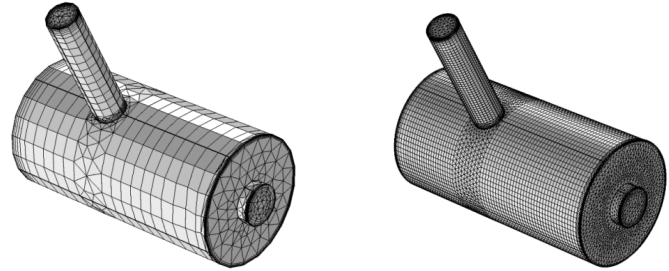

**Figure 2**: Position of the source term (red) within the resonator. Small variations of the radius and the height of the source term have no significant effect on the simulation results.

**Figure 3**: Coarsest and finest mesh (mesh 0 and mesh 4, respectively) generated for the studies.

**Table 2**: Properties of the generated meshes (DOF: degrees of freedom). The simulations were done on a 64-bit computer with a processor speed of 2.5 GHz and 32 GB RAM.

|  | Mesh 0 | Mesh 1 | Mesh 2 | Mesh 3 | Mesh 4 |
| --- | --- | --- | --- | --- | --- |
| **Mesh elements** | 12,354 | 22,451 | 41,409 | 79,267 | 199,264 |
| **DOF VT** | 266,860 | 476,980 | 875,804 | 1,804,987 | 3,722,294 |
| **DOF AME** | 64,589 | 115,474 | 212,041 | 437,186 | 901,678 |
| **AME solution time** | 5 minutes | 7 minutes | 11 minutes | 24 minutes | 52 minutes |
| **VT solution time** | - | 1 week | - | - | - |

The acoustic pressure at the microphone position is then calculated using Equation 2.

Five structured meshes with different mesh size were generated during the studies (Figure 3). They are used to simulate the PA signal of the resonator between 8 to 62 kHz with an increment of 10 Hz. The AME model does not require boundary layers, however, they were generated throughout the resonator since mesh 1 was also used for the VT model.

**VT model**
The model is simulated using the Thermoviscous Acoustics, Frequency Domain COMSOL Multiphysics® module. The walls of the resonator are set as sound hard (no-slip and isothermal boundary conditions). The no-slip condition for viscous fluids assumes that the fluid will have zero velocity relative to the boundary and the isothermal condition assumes that there are no temperature fluctuations at the boundary. This creates large thermal and viscous gradients at the walls of the resonator and hence the need for boundary layers.

Air was selected from the COMSOL Multiphysics® material database as the propagating fluid in the resonator and is set at 20°C and at static pressure of 1013 hPa. Unlike in the AME model, the fluid properties of the VT model are temperature dependent. The source term is defined just like in the AME model.

The number of degrees of freedom in the VT model is larger than in the AME model since it has more variables. This and the nonlinearity introduced by the temperature dependency of the fluid properties is the reason why the VT model is much slower and computationally more demanding. The VT model was simulated using mesh 1 due to its long computation time. The PA signal of the VT model was calculated between 8 to 62 kHz with an increment of 50 Hz.

**Results and discussion**
Figure 4 shows the frequency response of the resonator obtained with the AME model. The plots indicate convergence since for different meshes similar spectral features are produced.

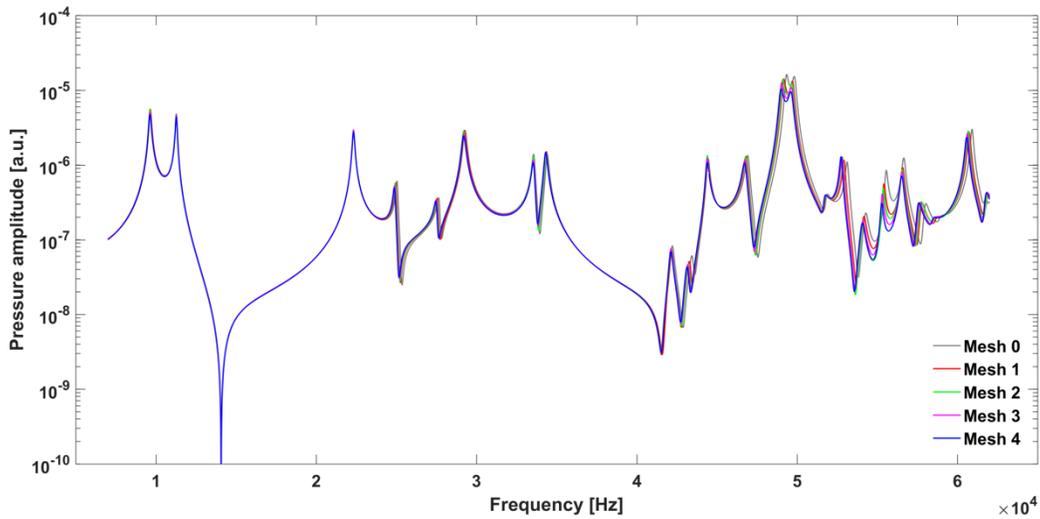

**Figure 4**: Frequency response plot of the AME model using the five generated meshes.

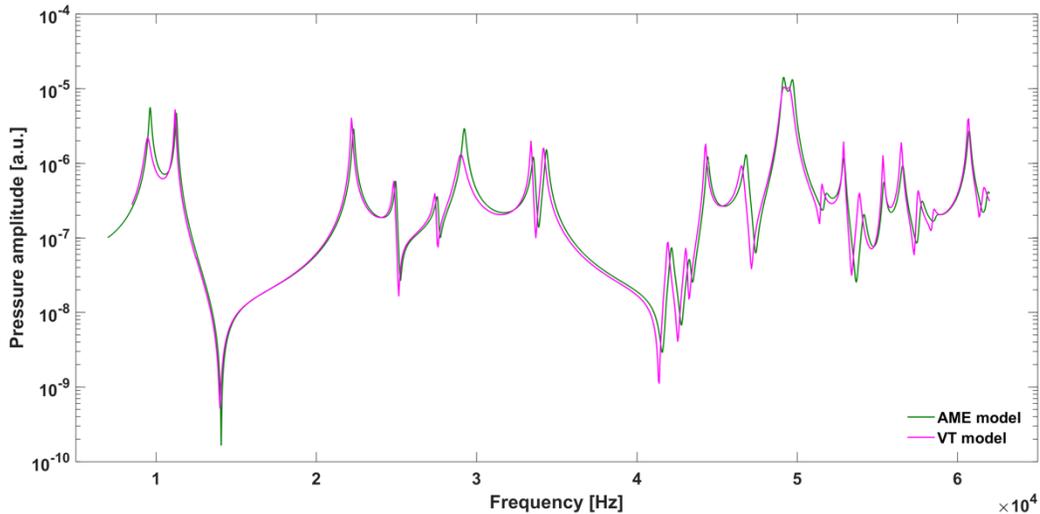

**Figure 5:** Frequency response curves of the AME model against the VT model. The response curves were obtained using mesh 1.

The simulation results of the AME and VT models are compared in Figure 5. All the resonances from the VT model are reproduced by the AME model while the resonance frequencies are reproduced with a deviation of less than 1.8%. Since in later investigations we are interested in a strong PA signal, we restrict the following discussion to the 14 resonances with amplitudes that exceed a value of $10^{-6}$.

At low modulation frequencies the resonances are attributed mainly to longitudinal modes. At higher frequencies (above 50 kHz) radial modes are also supported by the resonator thus accounting for the increase in the number of resonances within an interval of frequencies (Table 3). The broad resonance peak between 48 kHz and 51 kHz is a result of two overlapping resonance peaks at 49.200 kHz and 49.500 kHz.

It can be observed that the relative height of the resonance amplitudes corresponding to the two models depends on the main location of the mode within the resonator. If the mode is mainly located in the resonance cylinder where the surface area to volume ratio is large, the AME model underestimates the losses and has larger amplitudes than the VT model (ratio of AME to VT model resonance amplitude > 1).

If the mode is mainly located in the cavity cylinder where the surface area to volume ratio is small, the AME model slightly overestimates the losses and has smaller amplitudes than the VT model (resonance amplitude ratio <1). In some cases the mode occupies the cavity cylinder as well as the resonance cylinder. Then the amplitude ratio can be smaller or larger than 1. The relation between surface area to volume ratio and loss leads to the conclusion that the AME model tends to overestimate surface loss effects.

**Table 3**: Resonance frequency, corresponding mode, location of strong antinodes and amplitude ratio for the 14 strongest resonances. The location of the antinodes has been determined by lifting the lower limit of the depicted data range appropriately.

| $f_{res}$ in kHz | $|p|$-profile of acoustic mode | Main location of the mode | $\dfrac{A^{AME}(\omega_{res})}{A^{VT}(\omega_{res})}$ |
|---|---|---|---|
| 9.500 | 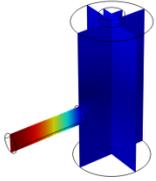 | 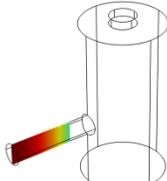 Resonance cylinder | 2.43 |
| 11.200 | 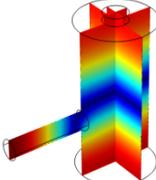 | 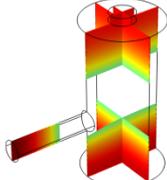 Resonance and cavity cylinder | 0.91 |
| 22.200 | 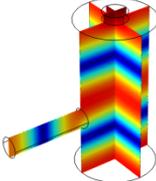 | 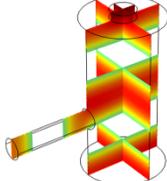 Resonance and cavity cylinder | 0.72 |
| 29.050 | 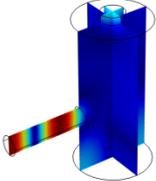 | 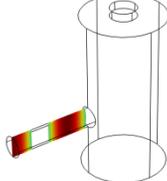 Resonance cylinder | 2.23 |
| 33.400 | 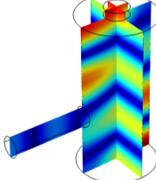 | 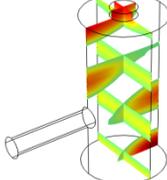 Cavity cylinder | 0.59 |
| 34.200 | 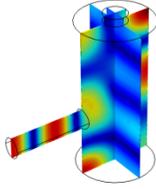 | 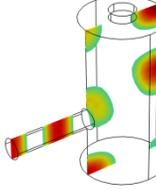<br>Resonance and cavity cylinder | 0.97 |
| 44.300 | 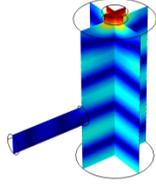 | 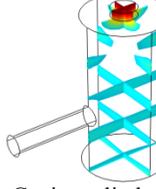<br>Cavity cylinder | 0.68 |
| 46.550 | 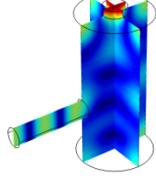 | 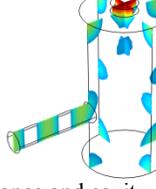<br>Resonance and cavity cylinder | 1.42 |
| 49.200 | 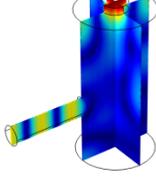 | 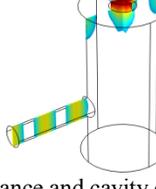<br>Resonance and cavity cylinder | 1.30 |
| 49.500 | 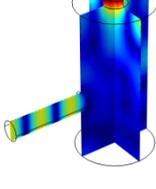 | 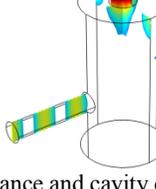<br>Resonance and cavity cylinder | 1.27 |
| 52.900 | 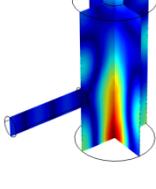 | 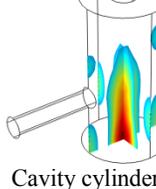<br>Cavity cylinder | 0.60 |
| 55.350 | 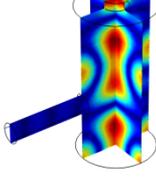 | 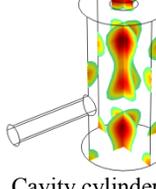<br>Cavity cylinder | 0.47 |
| 56.500 | 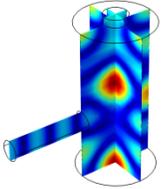 | 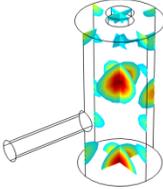  Cavity cylinder | 0.49 |
| 60.700 | 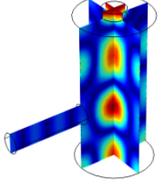 | 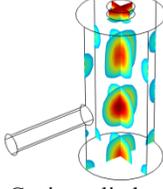  Cavity cylinder | 0.65 |

## Conclusion

The photoacoustic signal of a T-shaped resonator was determined using the viscothermal and the amplitude mode expansion model. Comparison of the results showed good accordance hence providing a much faster alternative for simulations of macroscopic photoacoustic resonators. Shape optimization of the resonator becomes a feasible option due to the gain in performance.